\documentclass[3p,11pt,authoryear]{elsarticle}

\usepackage{natbib}
\setcitestyle{square,numbers}
\usepackage{hyperref}
\hypersetup{
  colorlinks=true,
  linkcolor=black,
  citecolor=black,
  urlcolor=black
}
\makeatletter
\def\ps@pprintTitle{%
  \let\@oddhead\@empty
  \let\@evenhead\@empty
  \let\@oddfoot\@empty
  \let\@evenfoot\@empty
}
\makeatother

\usepackage{amssymb}

 \usepackage{lineno}
 \usepackage{etoolbox}
\AtBeginDocument{}
\usepackage{caption}
\begin{document}

\begin{frontmatter}



\title{Response to the $^7_\Lambda$He interpretation of \\MAMI's recent determination of $B_\Lambda(^3_\Lambda$H)}





\author{Ryoko~Kino\fnref{nishina,myfootnote}}
\author{Patrick~Achenbach\fnref{mainz}}
\author{Pascal~Klag\fnref{mainz}}
\author{Sho~Nagao\fnref{tokyo,qnsi}}
\author{\\Satoshi~N.~Nakamura\fnref{tokyo,qnsi,tohoku}}
\author{Kotaro~Nishi\fnref{tokyo}}
\author{Josef~Pochodzalla\fnref{mainz,him}}
\author{Tianhao~Shao\fnref{mainz}}
\author{\\ \it{on behalf of the A1 Collaboration}}

\affiliation[nishina]{
            organization={RIKEN Nishina Center for Accelerator-Based Science},
            city={Saitama},
            postcode={351-0198}, 
            country={Japan}}
\affiliation[mainz]{
            organization={Institute for Nuclear Physics, Johannes Gutenberg University Mainz},
            city={Mainz},
            postcode={55099}, 
            country={Germany}}
\affiliation[tokyo]{
            organization={Graduate School of Science, The University of Tokyo},
            city={Tokyo},
            postcode={113-0033}, 
            country={Japan}}
\affiliation[qnsi]{
            organization={Quark Nuclear Science Institute, Graduate School of Science, The University of Tokyo},
            city={Tokyo},
            postcode={113-0033}, 
            country={Japan}}
\affiliation[tohoku]{
            organization={Department of Physics, Graduate School of Science, Tohoku University},
            city={Miyagi},
            postcode={980-8578}, 
            country={Japan}}
\affiliation[him]{
            organization={Helmholtz Institute Mainz, GSI Helmholtzzentrum for Schwerionenforschung, Darmstadt, and Johannes Gutenberg University Mainz},
            city={Mainz},
            postcode={55099}, 
            country={Germany}}
           
 \fntext[myfootnote]{Corresponding author. E-mail: ryoko.kino@riken.jp}

\begin{abstract}
We respond to the recent suggestion by A. Gal~\citep{gal2026questioning} 
that the sharp pion-momentum peak at $p_{\pi^-} \approx 113.8$~MeV/$c$ 
observed in our $^7\mathrm{Li}(e,e^\prime K^+)$ electroproduction 
experiment at MAMI~\citep{kino2026precise} originates from 
$^7_\Lambda\mathrm{He}$ weak decay rather than from 
$^3_\Lambda\mathrm{H} \to \pi^- + {}^3\mathrm{He}$ as we reported. 
We present quantitative arguments against this interpretation and 
conclude that the $^3_\Lambda\mathrm{H}$ assignment remains the most 
well-supported interpretation of the data.
\end{abstract}



\end{frontmatter}



\section{Introduction}
\label{sec:Intro}

The precise determination of the $\Lambda$ binding energy of the hypertriton, $B_\Lambda(^3_\Lambda\mathrm{H})$, is a longstanding challenge in hypernuclear physics. Our recent measurement via decay pion spectroscopy at MAMI~\citep{kino2026precise} obtained $B_\Lambda(^3_\Lambda\mathrm{H}) = 0.523 \pm 0.013_\mathrm{stat} 
\pm 0.075_\mathrm{syst}$~MeV, significantly larger than the value inferred from earlier nuclear emulsion data. A.~Gal has suggested~\citep{gal2026questioning} that the sharp pion-momentum peak at $p_{\pi^-} \approx 113.8$~MeV/$c$ on which this result is based may instead originate from 
$^7_\Lambda\mathrm{He}(1/2^+_\mathrm{g.s.}) \to \pi^- + {}^7\mathrm{Li}(1/2^-, E_x = 478~\mathrm{keV})$ weak decay. 
This alternative interpretation would require $B_\Lambda(^7_\Lambda\mathrm{He}) = 5.84 \pm 0.07$~MeV, a value not directly measured but inferred from theoretical 
shell-model arguments applied to the $A=7$ isospin triplet~\citep{gal2026questioning}. 
This value is not only in tension with the direct spectroscopic measurement by the JLab HKS collaboration, $B_\Lambda(^7_\Lambda\mathrm{He}) = 5.55 \pm 0.10_\mathrm{stat} \pm 0.11_\mathrm{syst}$~MeV~\citep{gogami2016spectroscopy}, but also exceeds the world average of existing experimental data~\citep{eckert2021chart}.

We note that prior to the submission of Ref.~\citep{kino2026precise}, we had already carefully examined this alternative interpretation 
within our collaboration. We concluded that, while the $^7_\Lambda\mathrm{He}$ hypothesis cannot be excluded on the basis of the pion momentum alone, the $^3_\Lambda\mathrm{H}$ assignment is strongly favored by the data. In this Reply, we present these arguments in detail.

\section{Arguments against the $^7_\Lambda$He interpretation}

\subsection{Absence of the predicted companion peak}

If the sharp peak at $p_{\pi^-} \approx 113.8$~MeV/$c$ originates from $^7_\Lambda\mathrm{He}(1/2^+_\mathrm{g.s.}) \to \pi^- + {}^7\mathrm{Li}(1/2^-, E_x = 478~\mathrm{keV})$ weak decay as suggested in Ref.~\citep{gal2026questioning}, the ground-state transition $^7_\Lambda\mathrm{He}(1/2^+_\mathrm{g.s.}) \to \pi^- + {}^7\mathrm{Li}(3/2^-_\mathrm{g.s.})$ should also appear at $p_{\pi^-} \approx 114.5$~MeV/$c$. Indeed, Gal himself points out in the closing remark of Ref.~\citep{gal2026questioning} that such a line is expected with approximately twice the intensity of the $113.8$~MeV/$c$ line, based on the shell-model calculation of Ref.~\citep{gal2009pi}.

\begin{figure}[htbp]
\begin{center}
\includegraphics[width=0.7\linewidth]{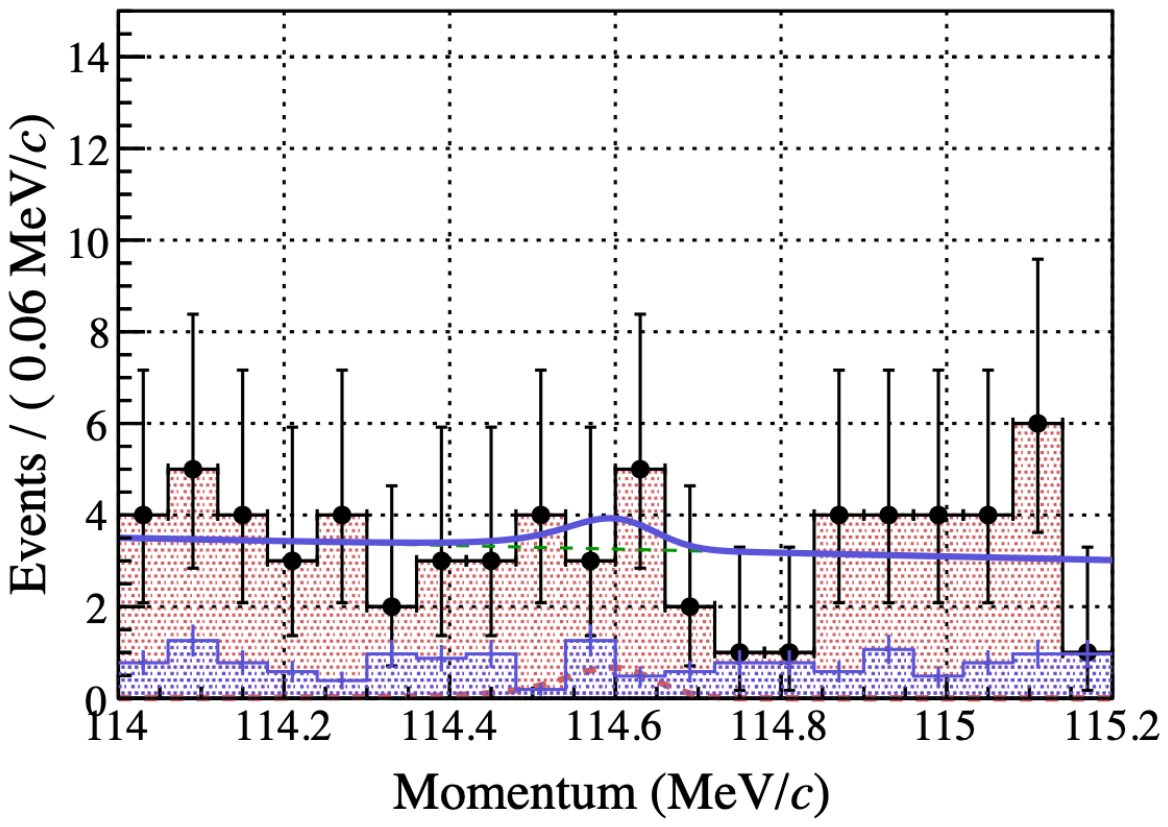}
\vspace{-30pt}
\caption[]{Measured decay $\pi^-$ momentum distribution near $p_{\pi^-} \approx 114.5$~MeV/$c$. 
The histograms show the spectra for true (red) and accidental (blue) coincidences, together with the result of an unbinned signal-plus-background fit using a Landau-Gaussian convolution function with shape parameters constrained by the $^4_\Lambda\mathrm{H}$ fit. 
No statistically significant excess is observed, and a profile-likelihood scan gives an upper limit of $N_\mathrm{UL} = 8.7$ events at 90\% confidence level.}
\label{fig:unbinnedfit}
\end{center}
\end{figure}

Figure~\ref{fig:unbinnedfit} shows the decay pion momentum spectrum near $p_{\pi^-} \approx 114.5$~MeV/$c$ together with the result of an unbinned signal-plus-background fit, where the signal component is modeled as a Landau-Gaussian convolution with shape parameters constrained by the $^4_\Lambda\mathrm{H}$ fit. No statistically significant excess is observed. A profile-likelihood scan of the signal yield gives an upper limit of $N_\mathrm{UL} = 8.7$ events at 90\% confidence level within $\pm3\sigma$ of the expected signal position $p_{\pi^-} \approx 114.5$~MeV/$c$.

Gal~\cite{gal2026questioning} predicts that if the $p_{\pi^-} \approx 113.8$~MeV/$c$ line arises from $^7_\Lambda\mathrm{He}(1/2^+_\mathrm{g.s.}) \to \pi^- + {}^7\mathrm{Li}(1/2^-,\, E_x = 0.478~\mathrm{MeV})$ weak decay, a companion line near $p_{\pi^-} \approx 114.5$~MeV/$c$ from $^7_\Lambda\mathrm{He}(1/2^+_\mathrm{g.s.}) \to \pi^- + {}^7\mathrm{Li}(3/2^-_\mathrm{g.s.})$ should appear with approximately twice the intensity. The signal yield of the $113.8$~MeV/$c$ line determined from our fit is $S=17.8$~events within $\pm3\sigma$, implying an expected yield of $2S \approx 36$~events near 114.5~MeV/$c$. This expectation exceeds our upper limit by more than a factor of four, providing a direct quantitative argument against the $^7_\Lambda\mathrm{He}$ interpretation of the 113.8~MeV/$c$ line.

Turning Gal's argument around, the upper limit at 114.5~MeV/$c$ can itself be used to constrain the possible contribution of $^7_\Lambda\mathrm{He}$ to the 113.8~MeV/$c$ line. Applying the predicted intensity ratio of approximately 2:1, the upper limit of $N_\mathrm{UL} = 8.7$~events at 114.5~MeV/$c$ implies that at most $N_\mathrm{UL}/2 \approx 4.4$~events in the 113.8~MeV/$c$ peak could be attributed to $^7_\Lambda\mathrm{He}$ decay. Since the observed yield at 113.8~MeV/$c$ is $S \sim 17.8$~events, the $^7_\Lambda\mathrm{He}$ hypothesis can account for at most $ \sim$25\% of the observed signal, even under the most favorable assumptions. The remaining $\sim$75\% of the peak must then be attributed to another source, for which $^3_\Lambda\mathrm{H}$ is the viable candidate given the observed pion momentum.

\subsection{Inconsistency with the direct spectroscopic measurement of $B_\Lambda(^7_\Lambda\mathrm{He})$}

The $^7_\Lambda\mathrm{He}$ interpretation requires $B_\Lambda(^7_\Lambda\mathrm{He}) = 5.84 \pm 0.07$~MeV, which Gal shows to be consistent with a value inferred indirectly from the $A=7$ isospin triplet using $B_\Lambda(^7_\Lambda\mathrm{Li})$ from KEK-SKS and DA$\Phi$NE-FINUDA measurements. While we acknowledge this internal consistency, we note a fundamental distinction between these indirect estimates and the direct spectroscopic measurement by the JLab HKS collaboration, which obtained $B_\Lambda(^7_\Lambda\mathrm{He}) = 5.55 \pm 0.10_\mathrm{stat} \pm 0.11_\mathrm{syst}$~MeV~\citep{gogami2016spectroscopy} via the missing-mass method in $^7\mathrm{Li}(e,e^\prime K^+)$ electroproduction. In this measurement, the momentum scale was calibrated using the well-known masses of $\Lambda$ and $\Sigma^0$ produced from a hydrogen target, providing an absolute and model-independent reference free from nuclear structure uncertainties. The $^7_\Lambda\mathrm{He}$ interpretation requires this direct measurement to be in error by approximately $2\sigma$, with no experimental justification for such a discrepancy. We therefore regard the JLab HKS result as the more reliable constraint on $B_\Lambda(^7_\Lambda\mathrm{He})$.

\section{Support for the $^3_\Lambda\mathrm{H}$ interpretation}

\subsection{The significance of the deviation from emulsion data}

Gal argues that our value of $B_\Lambda(^3_\Lambda\mathrm{H}) = 0.523 \pm 0.013_\mathrm{stat} \pm 0.075_\mathrm{syst}$~MeV deviates by more than $4\sigma$ from the weighted average of preceding measurements~\citep{eckert2021chart}. However, this average is dominated by nuclear emulsion data, and the systematic uncertainties inherent to emulsion analysis of the hypertriton are particularly difficult to quantify. As discussed in Ref.~\citep{kino2026precise}, the apparent systematic differences of approximately 0.2~MeV between binding energies evaluated from two- and three-body decay modes in emulsion data may be attributed to uncertainties in the range-energy relation used in emulsion analysis~\citep{bohm1970investigation}. No reliable systematic uncertainty has been assigned to these earlier emulsion results. We therefore caution against treating the emulsion-based weighted average as a firm reference value for the hypertriton binding energy.

\subsection{Consistency with independent experimental results}

Furthermore, our value of $B_\Lambda(^3_\Lambda\mathrm{H})$ is consistent with the result reported by the STAR Collaboration, $B_\Lambda = 0.406 \pm 0.120_\mathrm{stat} \pm 0.110_\mathrm{syst}$~MeV~\citep{STAR2020}, which was obtained independently using an entirely different experimental technique. The STAR result itself also deviates significantly from the emulsion average, suggesting that the discrepancy may reflect unquantified systematic uncertainties in the emulsion data rather than an error in the more recent measurements. While we acknowledge that the ALICE Collaboration reported a value consistent with the emulsion average~\citep{Acharya2023}, the current spread among experimental results underscores the difficulty of this measurement and the importance of direct decay pion spectroscopy as employed in the present work.


\section{Summary}
\label{sec:summary}

We have presented quantitative arguments against the 
$^7_\Lambda\mathrm{He}$ interpretation proposed in 
Ref.~\citep{gal2026questioning}. The most direct evidence 
is the absence of the companion peak at 
$p_{\pi^-} \approx 114.5$~MeV/$c$: applying Gal's own 
predicted intensity ratio of 2:1, the upper limit of 
$N_\mathrm{UL} = 8.7$ events at this position implies 
that $^7_\Lambda\mathrm{He}$ decay can account for at most 
$\sim$25\% of the observed signal at 113.8~MeV/$c$, 
leaving the majority of the peak unexplained under the 
$^7_\Lambda\mathrm{He}$ hypothesis. Furthermore, this 
interpretation requires the direct spectroscopic 
measurement of $B_\Lambda(^7_\Lambda\mathrm{He})$ by the 
JLab HKS experiment to be significantly in error, for 
which there is no experimental justification. Our value 
of $B_\Lambda(^3_\Lambda\mathrm{H})$ is consistent with 
the STAR result obtained by an independent method, 
and the deviation from the emulsion average can be 
understood in terms of unquantified systematic 
uncertainties in those earlier measurements. We therefore 
conclude that the assignment of the 
$p_{\pi^-} \approx 113.8$~MeV/$c$ peak to 
$^3_\Lambda\mathrm{H} \to \pi^- + {}^3\mathrm{He}$ weak 
decay, as reported in Ref.~\citep{kino2026precise}, 
remains the most well-supported interpretation of the 
data. A definitive resolution will require future 
dedicated experiments with higher statistics.









\end{document}